\documentstyle[12pt,epsfig]{article}
\advance\textheight by \topskip
\begin{document}
\tolerance=100000
\thispagestyle{empty}
\setcounter{page}{0}

\newcommand{\be}{\begin{equation}}
\newcommand{\ee}{\end{equation}}
\newcommand{\br}{\begin{eqnarray}}
\newcommand{\er}{\end{eqnarray}}
\newcommand{\ba}{\begin{array}}
\newcommand{\ea}{\end{array}}
\newcommand{\bi}{\begin{itemize}}
\newcommand{\ei}{\end{itemize}}
\newcommand{\bn}{\begin{enumerate}}
\newcommand{\en}{\end{enumerate}}
\newcommand{\bc}{\begin{center}}
\newcommand{\ec}{\end{center}}
\newcommand{\ul}{\underline}
\newcommand{\ol}{\overline}
\newcommand{\ar}{\rightarrow}
\newcommand{\sm}{${\cal {SM}}$}
\newcommand{\susy}{{{SUSY}}}
\newcommand{\Dir}{\kern -6.4pt\Big{/}}
\newcommand{\Dirin}{\kern -12.4pt\Big{/}\kern 4.4pt}
\newcommand{\DGir}{\kern -6.0pt\Big{/}}
\def\pp{\ifmmode{pp} \else{$pp$} \fi}
\def\CC{\ifmmode{{\it C.C.}} \else{$\mbox{\it C.C.}$} \fi}
\newcommand{\jet}{\ifmmode{{\mathrm{j}}} 
                   \else{${\mathrm{j}}$}\fi}
\newcommand{\jj}{\ifmmode{{\mathrm{2~jets}}} 
                   \else{${\mathrm{2~jets}}$}\fi}
\newcommand{\bqbqH}{\ifmmode{bU\ar bDH^+} 
                   \else{$bU\ar bDH^+$}\fi}
\newcommand{\bqbqW}{\ifmmode{bU\ar bDW^+} 
                   \else{$bU\ar bDW^+$}\fi}
\newcommand{\qqbbW}{\ifmmode{U\bar D\ar b\bar bW^+} 
                   \else{$U\bar D\ar b\bar bW^+$}\fi}
\newcommand{\bqbqtn}{\ifmmode{bU\ar bD\tau^+\nu_\tau} 
                   \else{$bU\ar bD\tau^+\nu_\tau$}\fi}
\newcommand{\qqbbtn}{\ifmmode{U\bar D\ar b\bar b\tau^+\nu_\tau} 
                   \else{$U\bar D\ar b\bar b\tau^+\nu_\tau$}\fi}
\newcommand{\Htn}{\ifmmode{H^\pm\ar \tau\nu_\tau}
                   \else{$H^\pm\ar \tau\nu_\tau$}\fi}
\newcommand{\Wtn}{\ifmmode{W^\pm\ar \tau\nu_\tau}
                   \else{$W^\pm\ar \tau\nu_\tau$}\fi}
\def\mssm{\ifmmode{{\cal {MSSM}}}\else{${\cal {MSSM}}$}\fi}
\def\bbtowh{\ifmmode{b\bar b\ar W^\pm H^\mp}\else{${b\bar b\ar W^\pm H^\mp}$}\fi}
\def\ggtowh{\ifmmode{gg\ar W^\pm H^\mp}\else{${gg\ar W^\pm H^\mp}$}\fi}
\def\wpmhmp{\ifmmode{W^\pm H^\mp}\else{${W^\pm H^\mp}$}\fi}
\def\MH{\ifmmode{{M_{H}}}\else{${M_{H}}$}\fi}
\def\Mh{\ifmmode{{M_{h}}}\else{${M_{h}}$}\fi}
\def\MA{\ifmmode{{M_{A}}}\else{${M_{A}}$}\fi}
\def\MHpm{\ifmmode{{M_{H^\pm}}}\else{${M_{H^\pm}}$}\fi}
\def\Hpm{\ifmmode{{{H^\pm}}}\else{${{H^\pm}}$}\fi}
\def\tb{\ifmmode{\tan\beta}\else{$\tan\beta$}\fi}
\def\ctb{\ifmmode{\cot\beta}\else{$\cot\beta$}\fi}
\def\ta{\ifmmode{\tan\alpha}\else{$\tan\alpha$}\fi}
\def\cta{\ifmmode{\cot\alpha}\else{$\cot\alpha$}\fi}
\def\tba{\ifmmode{\tan\beta=1.5}\else{$\tan\beta=1.5$}\fi}
\def\tbb{\ifmmode{\tan\beta=30}\else{$\tan\beta=30.$}\fi}
\def\cab{\ifmmode{c_{\alpha\beta}}\else{$c_{\alpha\beta}$}\fi}
\def\sab{\ifmmode{s_{\alpha\beta}}\else{$s_{\alpha\beta}$}\fi}
\def\cba{\ifmmode{c_{\beta\alpha}}\else{$c_{\beta\alpha}$}\fi}
\def\sba{\ifmmode{s_{\beta\alpha}}\else{$s_{\beta\alpha}$}\fi}
\def\ca{\ifmmode{c_{\alpha}}\else{$c_{\alpha}$}\fi}
\def\sa{\ifmmode{s_{\alpha}}\else{$s_{\alpha}$}\fi}
\def\cb{\ifmmode{c_{\beta}}\else{$c_{\beta}$}\fi}
\def\sb{\ifmmode{s_{\beta}}\else{$s_{\beta}$}\fi}

\def\Ord{\buildrel{\scriptscriptstyle <}\over{\scriptscriptstyle\sim}}
\def\OOrd{\buildrel{\scriptscriptstyle >}\over{\scriptscriptstyle\sim}}
\def\pl #1 #2 #3 {{\it Phys.~Lett.} {\bf#1} (#2) #3}
\def\np #1 #2 #3 {{\it Nucl.~Phys.} {\bf#1} (#2) #3}
\def\zp #1 #2 #3 {{\it Z.~Phys.} {\bf#1} (#2) #3}
\def\pr #1 #2 #3 {{\it Phys.~Rev.} {\bf#1} (#2) #3}
\def\prep #1 #2 #3 {{\it Phys.~Rep.} {\bf#1} (#2) #3}
\def\prl #1 #2 #3 {{\it Phys.~Rev.~Lett.} {\bf#1} (#2) #3}
\def\mpl #1 #2 #3 {{\it Mod.~Phys.~Lett.} {\bf#1} (#2) #3}
\def\rmp #1 #2 #3 {{\it Rev. Mod. Phys.} {\bf#1} (#2) #3}
\def\sjnp #1 #2 #3 {{\it Sov. J. Nucl. Phys.} {\bf#1} (#2) #3}
\def\cpc #1 #2 #3 {{\it Comp. Phys. Comm.} {\bf#1} (#2) #3}
\def\xx #1 #2 #3 {{\bf#1}, (#2) #3}
\def\preprint{{\it preprint}}

\begin{flushright}
{RAL-TR-2000-005}\\ 
{March 2000\hspace*{.5 truecm}}\\
\end{flushright}

\vspace*{\fill}

\begin{center}
{\Large \bf 
The $W^\pm h$ decay channel\\[0.25cm]
as a probe of charged Higgs boson production\\[0.45cm]
at the Large Hadron Collider}\\[1.5cm]
{\large Stefano 
Moretti\footnote{Electronic mail: moretti@v2.rl.ac.uk}}\\[0.4 cm]
{\it Rutherford Appleton Laboratory,}\\
{\it Chilton, Didcot, Oxon OX11 0QX, UK.}\\[0.75cm]
\end{center}
\vspace*{\fill}

\begin{abstract}
{\noindent
We analyse the chances of detecting charged Higgs bosons of the
Minimal Supersymmetric Standard Model (MSSM) at the Large Hadron Collider
(LHC) in the $W^\pm h$ mode, followed by the dominant decay of
the lightest Higgs scalar, $h\to b\bar b$. 
If the actual value of $M_h$ is
already known, this channel offers possibly the optimal final
state kinematics for charged Higgs discovery, thanks to the narrow
resonances appearing around the $W^\pm$ and $h$ masses. 
Besides, within the MSSM, the 
$H^\pm\to W^\pm h$ decay rate is significant for not too
large  $\tan\beta$ values, thus offering the possibility of accessing
a region of MSSM parameter space left uncovered by  other search channels.
We consider both strong (QCD)
 and electroweak (EW) `irreducible' backgrounds 
in the $3b$-tagged channel to the $gg\to t\bar
bH^-$ production process that had
not been taken into account in previous analyses.
After a series of kinematic cuts, the largest of these processes is  $t\bar
bW^\pm h$ production in the
continuum. However, for optimum $\tan\beta$, i.e., between 2 and 3, the 
charged Higgs boson signal overcomes this background and a narrow
discovery region survives around $M_{H^\pm}\approx200$ GeV.
 }
\end{abstract}

\vspace*{\fill}
\newpage

\section*{1. Introduction}

 The discovery of  charged Higgs bosons \cite{HHG} will
provide a concrete evidence of the multi-doublet structure of the Higgs
sector. Recent efforts have focused on their relevance to Supersymmetry (SUSY),
in particular in the MSSM, which incorporates
exactly two Higgs doublets, yielding -- after spontaneous EW 
symmetry breaking -- five physical Higgs states:  
the neutral pseudoscalar ($A$), the lightest ($h$)
and heaviest ($H$) neutral scalars and two charged ones ($H^\pm$).

 In much of the parameter space preferred by SUSY, namely
$M_{H^\pm}\ge {M_{W^\pm}}$ and $1<\tan\beta<m_t/m_b$
 \cite{CMS,ATLAS}, the LHC
 will provide the greatest opportunity for the 
discovery of $H^\pm$ particles. 
In fact, over the above $\tan\beta$ region, the Tevatron
(Run 2) discovery potential is limited to charged Higgs masses smaller than 
$m_t$ \cite{FNAL}. 

 However, at the LHC, whereas the detection of light charged Higgs
bosons (with $M_{H^\pm}<m_t$) is rather straightforward in the decay channel 
$t\to bH^+$ for most $\tan\beta$ values, 
thanks to the huge top-antitop production rate,   
the search is notoriously difficult for
heavy  masses (when $M_{H^\pm}>m_t$), because of the large reducible and
irreducible backgrounds associated with the main decay mode $H^-\to b\bar t$, 
following the dominant production channel $bg\to t H^-$ \cite{bg}.
(Notice that the rate of the latter exceeds by far other possible
production modes \cite{bq}--\cite{ioekosuke}, this rendering it  the
only viable channel at the CERN machine in the heavy mass region.)

 The analysis of the $H^-\to b\bar t$ signature 
has been the subject of many debates
\cite{roger}--\cite{roy1}, whose conclusion is that  the LHC discovery
potential is satisfactory, 
but only provided that $\tan\beta$ is small ($\Ord1.5$) or large
($\OOrd30$)
 enough and the charged Higgs boson mass is below 600 GeV or so.

 A recent analysis \cite{kosuke} has shown that the $\tau\nu$ decay mode,
indeed dominant for light charged Higgs states and exploitable below
the top threshold for any accessible $\tan\beta$ \cite{ray},
can be used at the LHC
even in the large $M_{H^\pm}$ case, in order to discover $H^\pm$ scalars in the
parameter range $\tan\beta\OOrd3$ and 200 GeV $<M_{H^\pm}<1$ TeV. Besides,
if the distinctive $\tau$ polarisation \cite{Ben} is used
in this channel, the latter can provide at least as good a heavy
$H^\pm$ signature as the $H^-\to b\bar t$ decay mode (for the large 
$\tan\beta$ regime \cite{newroy,work}).

\begin{figure}[p]
~\epsfig{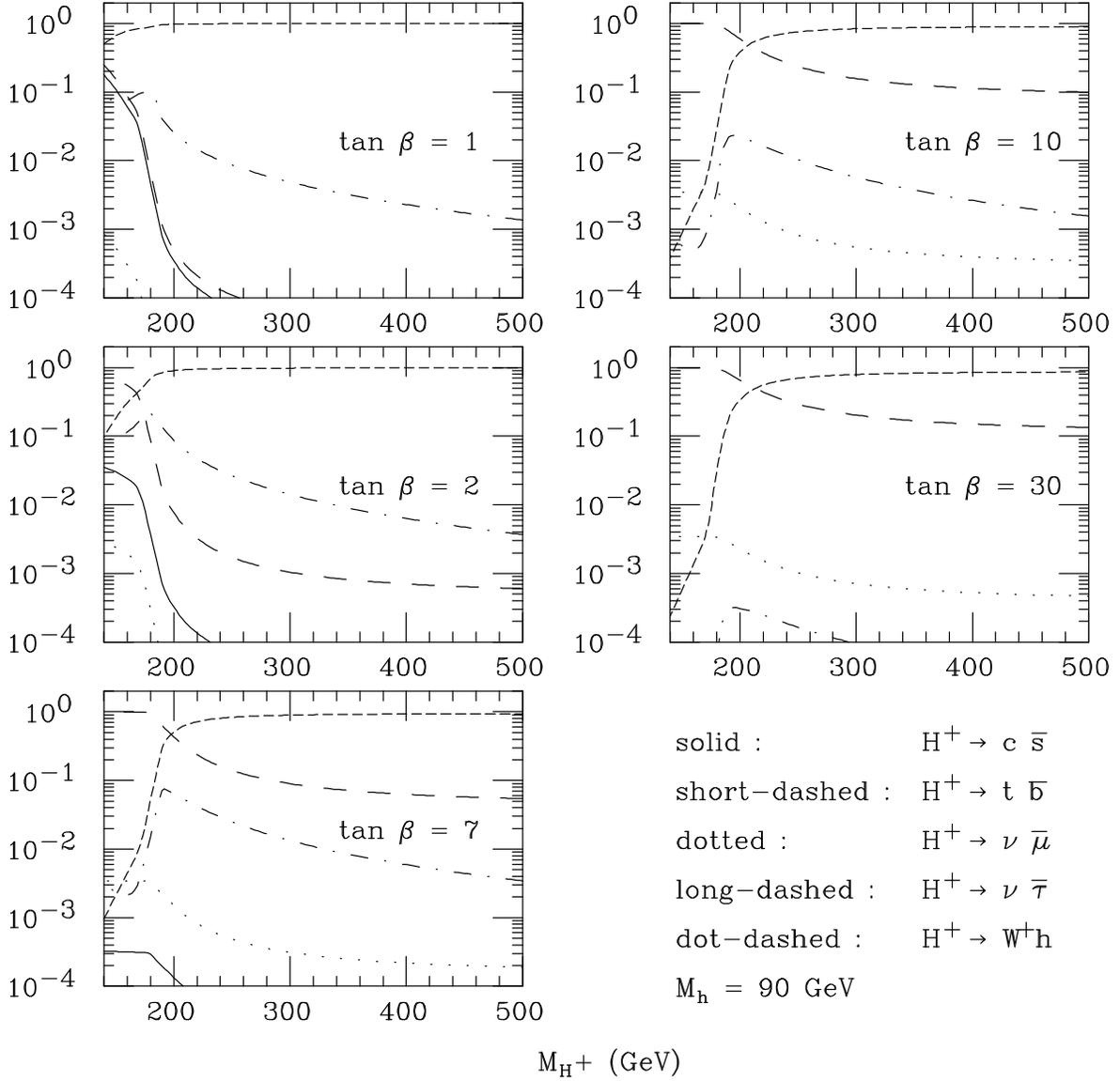}
\vspace*{0.25cm}
\caption{Dominant branching ratios of the charged Higgs
boson of the MSSM for selected values of $\tan\beta$ over the
mass range 140 GeV $\Ord M_{H^\pm}\Ord$ 500 GeV.
The mass of the lightest Higgs boson has been fixed at $M_h=90$ GeV.}
\label{fig:BRs}
\end{figure}

At present then, it is the $\tan\beta\Ord3$ region of the MSSM
which ought to be explored through other decay modes, 
especially those where direct mass
reconstruction is possible. The most obvious of these is the 
$H^\pm\to W^{\pm(*)} h$ channel  \cite{BR1} (see also \cite{BR2}), 
proceeding via the 
production of a charged gauge boson and the lightest Higgs scalar
of the MSSM, with the former on- or off-shell depending on  the
relative values of $M_{H^\pm}$ and $M_{h}$. In fact, its branching ratio 
(BR) can be rather large,
competing with the bottom-top decay mode and overwhelming 
the tau-neutrino one for $M_{H^\pm}\OOrd m_t$  
at low $\tan\beta$: see Figs.~\ref{fig:BRs}--\ref{fig:BRh}.
Besides, under the assumption that the $h$ scalar has previously
been discovered (which we embrace here), its kinematics is
rather constrained, around two resonant decay modes, $W^\pm\to$ 2 jets
(or lepton-neutrino) and $h\to b\bar b$, an aspect which allows for 
a significant reduction of the QCD background.
As demonstrated in Ref.~\cite{whroy}, signals of charged Higgs bosons
in the $2\Ord\tan\beta\Ord3$ range can be seen in this channel, provided
that 200 GeV $\Ord M_{H^\pm}\Ord220$ GeV 
(see also \cite{whketevi} for an experimental
simulation). The above lower limit on $\tan\beta$ corresponds
 to the border of the exclusion region drawn from LEP2 direct searches
for the MSSM $h$ scalar, whose mass bound is now set at $M_h\OOrd100$ GeV or so
\cite{mh}.

It is the purpose of this letter
that of resuming the studies of Ref.~\cite{whroy}, 
by analysing the contribution to 
the background due to several irreducible processes, not considered
there, whose presence could spoil the feasibility
of charged Higgs searches in the $W^{\pm(*)} h$ mode of the MSSM. 

The plan of this paper is as follows. In the next Section we
discuss possible signals and backgrounds, their implementation
and list the values adopted for the various parameters needed
for their computation.
Section 3 is devoted to the presentation and discussion of the results.
Conclusions are in Section 4. 

\section*{2. Signals and backgrounds} 

We generate the signal cross sections by using the formulae of
Ref.~\cite{roy}. That is, we implement the $2\to3$ matrix element (ME)
for the process
\begin{equation}\label{signalME}
gg\to t\bar b H^- +~{\mathrm{charge~conjugate~(c.c.)}}.
\end{equation}
This nicely embeds both the $gg\to t\bar t\to t\bar b H^- +~{\mathrm{c.c.}}$
subprocess of top-antitop production and decay, which is dominant
for $m_t\OOrd M_{H^\pm}$, as well as the $bg\to tH^-$ + c.c. one of
$b\bar t$-fusion and $H^\pm$-bremsstrahlung, 
which is responsible for charged Higgs production
in the case $m_t\Ord M_{H^\pm}$ \cite{gg}.
The ME of process (\ref{signalME})
has been computed by means of the spinor 
techniques of 
Refs.~\cite{KS}--\cite{ioPRD}. 

In the $H^-\to W^{-(*)}h\to W^{-(*)} b\bar b$ channel, assuming high efficiency
and purity in selecting/rejecting $b$-/non-$b$-jets, possible 
irreducible background
processes are the following (we consider only the $gg$-initiated channels):
\begin{enumerate}
\item  the $t\bar b W^- h$ continuum;
\item  $t\bar b W^- Z$ production, especially when $M_Z\approx M_{h}$;
\item  the QCD induced case $t\bar b W^- g$;
\item  and, finally, $t\bar b W^- H$ and $t\bar b W^- A$ intermediate states;
\end{enumerate}
in which $H,h,A,Z,g\to b\bar b$, plus their c.c. channels.
Once the top quark appearing in the  above reactions decays, two
$W^\pm$ bosons are present in each event. We will eventually assume the 
$W^+W^-$ pair to
decay semi-leptonically to light-quark jets, electrons/muons and 
corresponding neutrinos. Furthermore, we will require to tag exactly
three $b$-jets in the final state (e.g., by using $\mu$-vertex or
high $p_T$ lepton techniques). The same `signature' was considered
in Ref.~\cite{whroy}, where only the `intrinsic' 
 $t\bar b H^-\to t\bar t b\bar b$ background and the QCD 
noise due to `$t\bar t$ + jet' events
were studied (with jet signifying here either a $b$-,
light-quark or gluon jet, the latter two mistagged for the former).

Both signal and background MEs have been integrated numerically by means of 
{\tt VEGAS} \cite{VEGAS} and, for test purposes, of {\tt RAMBO} 
\cite{RAMBO} and Metropolis \cite{hamid} as well.
While proceeding to the phase space integration, one also has to fold in the 
$(x,Q^2)$-dependent Parton Distribution Functions (PDFs)
 for the  two incoming gluons. These have 
been evaluated at leading-order, by means of the package 
MRS-LO(05A) \cite{MRS98LO}. 

The numerical values of the SM parameters are ($\ell=e,\mu$):
$$m_\ell=m_{\nu_\ell}=m_u=m_d=m_s=m_c=0,
$$
$$\qquad m_b=4.25~{\mathrm{GeV}},
\qquad m_t=175~{\mathrm{GeV}},$$
$$M_Z=91.2~{\mathrm {GeV}},\quad\quad \Gamma_Z=2.5~{\mathrm {GeV}},$$
\begin{equation}\label{param}
M_{W^\pm}=80.2~{\mathrm {GeV}},\quad\quad \Gamma_W=2.2~{\mathrm {GeV}}.
\end{equation}
As for the top width $\Gamma_t$, we have used the LO value calculated
within the MSSM (i.e., $\Gamma_t=1.55$ GeV if $M_{H^\pm} \gg m_t$).

Concerning the MSSM parameters, we  proceed as follows. For a start,
we assume that the mass of the lightest neutral Higgs particle (but not 
$\tan\beta$)
is already known, thanks to its discovery at either LEP2, Tevatron
(Run 2) or from early analyses at the LHC itself. Thus, for us,
$\Mh$ is a fixed parameter, assuming for reference the following discrete
values: e.g., 90, 100, 110, 120 and 130 GeV. Then we express
all other Higgs masses as a function of $M_{H^\pm}$ and $\tan\beta$.
For the pseudoscalar Higgs boson mass, the tree-level relation
$M_{H^\pm}^2=M_{W^\pm}^2+M_A^2$ is assumed. Radiative corrections then, of
arbitrary perturbative order, are in practice embedded in the $H$ mass and the
mixing angle $\alpha$. 
In general, notice that, at the `Renormalisation
Group improved' one-loop level  \cite{carena}, it is
only for very large values of the lightest stop mass and of the squark
mixing parameters that $M_h$ can escape the LEP2 bound in the low $\tan\beta$
region, on which we will focus most of our attention.

Finally, notice that we develop our discussion at the parton level, without 
considering fragmentation and 
hadronisation effects. Thus, jets are identified with the
partons from which they originate and all cuts are applied directly to the
latter. In particular, when selecting $b$-jets, a vertex tagging is implied,
with a finite efficiency, $\epsilon_b$, per each tag. 
Moreover, we assume no correlations among multiple tags, nor do we 
include misidentification of light-quark (including $c$-quark-)jets
produced in $W^\pm$ decays as $b$-jets. 

\section*{3. Results and discussion}

\begin{figure}[p]
~\epsfig{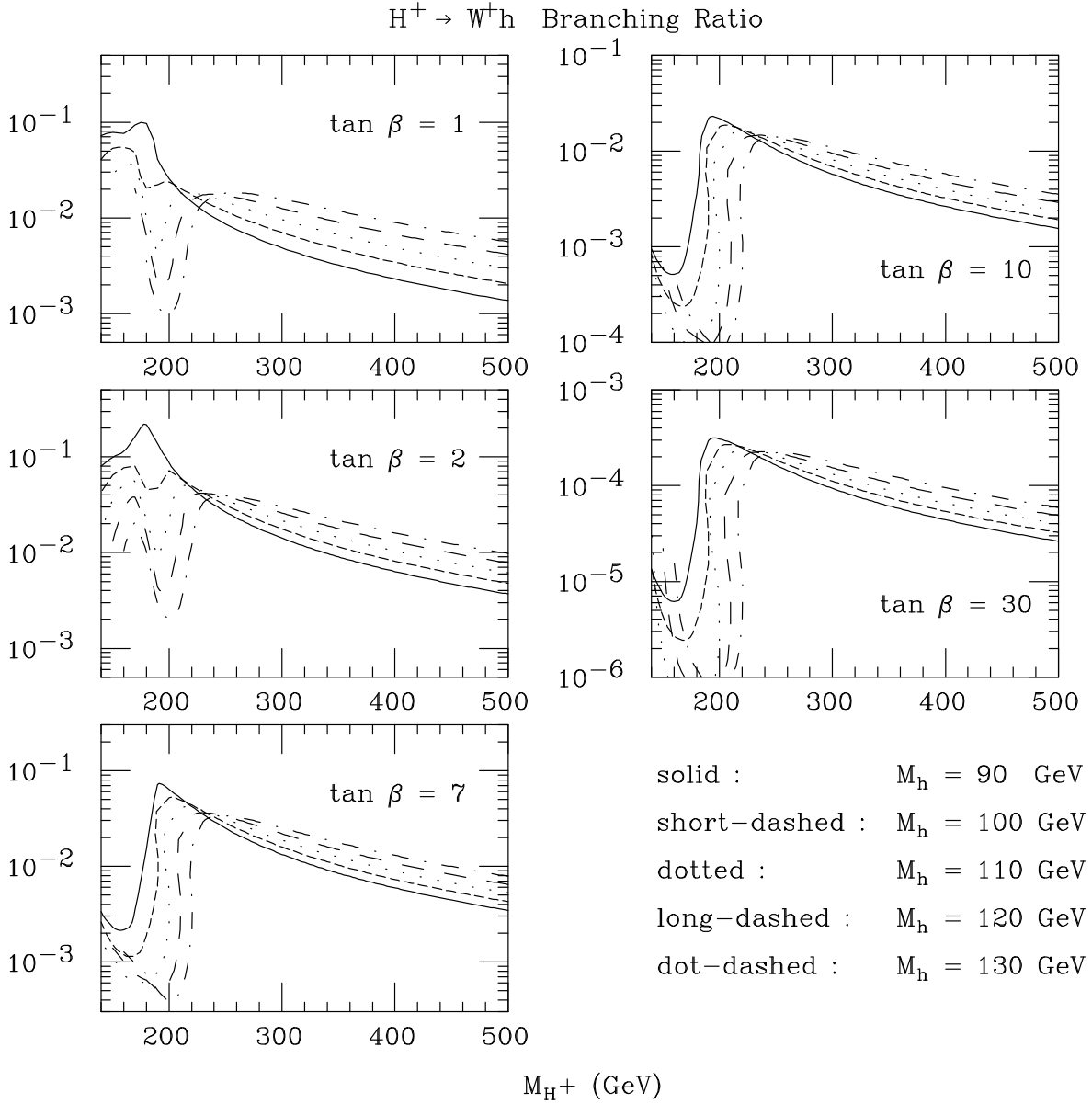}
\vspace*{0.25cm}
\caption{Branching ratios into $W^\pm h$ pairs of the charged Higgs
boson of the MSSM for selected values of $\tan\beta$ over the
mass range 140 GeV $\Ord M_{H^\pm}\Ord$ 500 GeV.
The mass of the lightest Higgs boson has been fixed at $M_h=90,100,110,120$
and 130 GeV.}
\label{fig:BRh}
\end{figure}

\begin{figure}[p]
~\epsfig{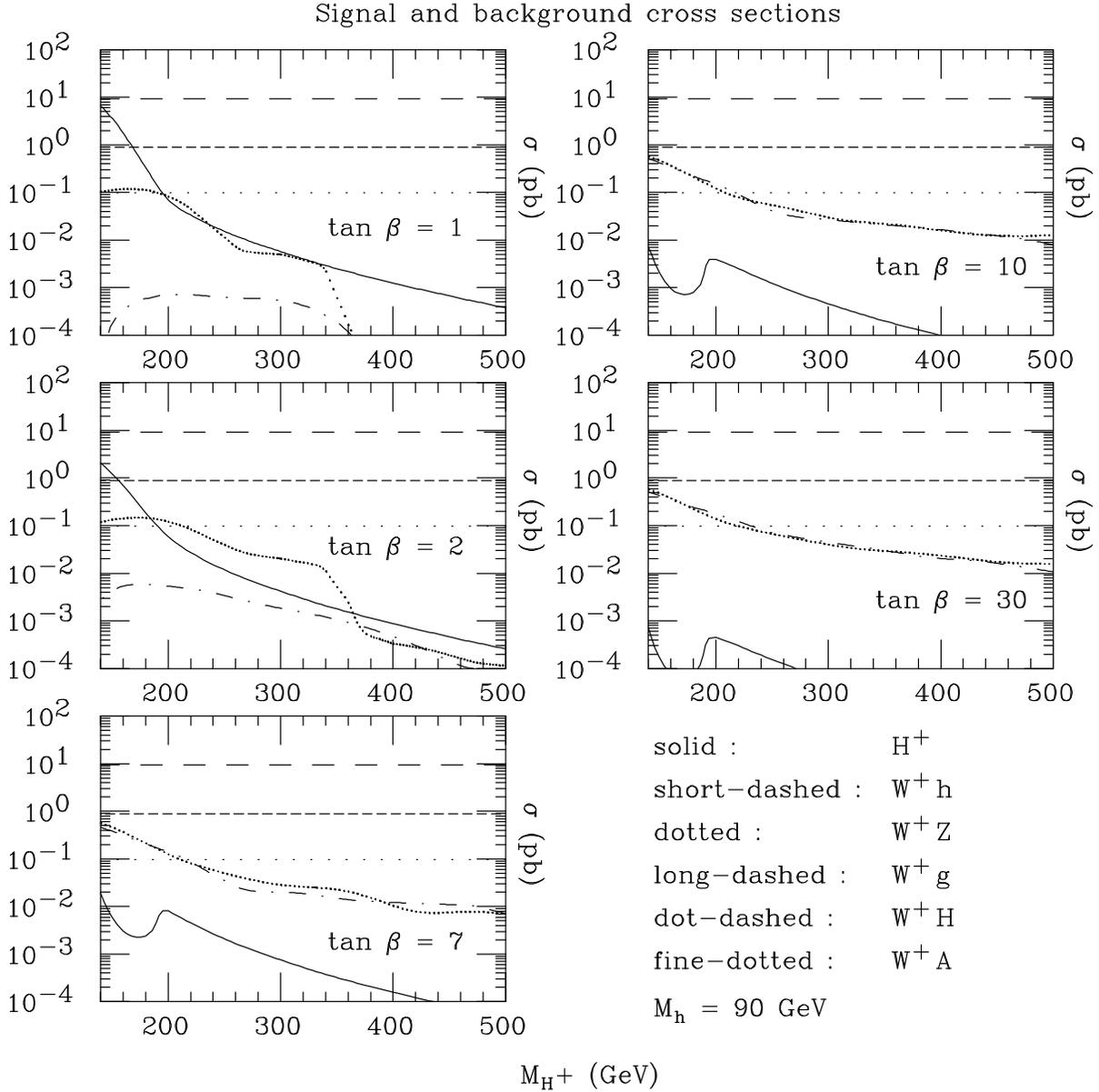}
\vspace*{0.25cm}
\caption{Cross sections of signal and backgrounds 
for selected values of $\tan\beta$ over the
mass range 140 GeV $\Ord M_{H^\pm}\Ord$ 500 GeV.
The mass of the lightest Higgs boson has been fixed at $M_h=90$ GeV.
Here, both the top quark and the $W^\pm$ boson are kept on-shell and
no decay rates and cuts are applied.}
\label{fig:cross90}
\end{figure}

\begin{figure}[p]
~\epsfig{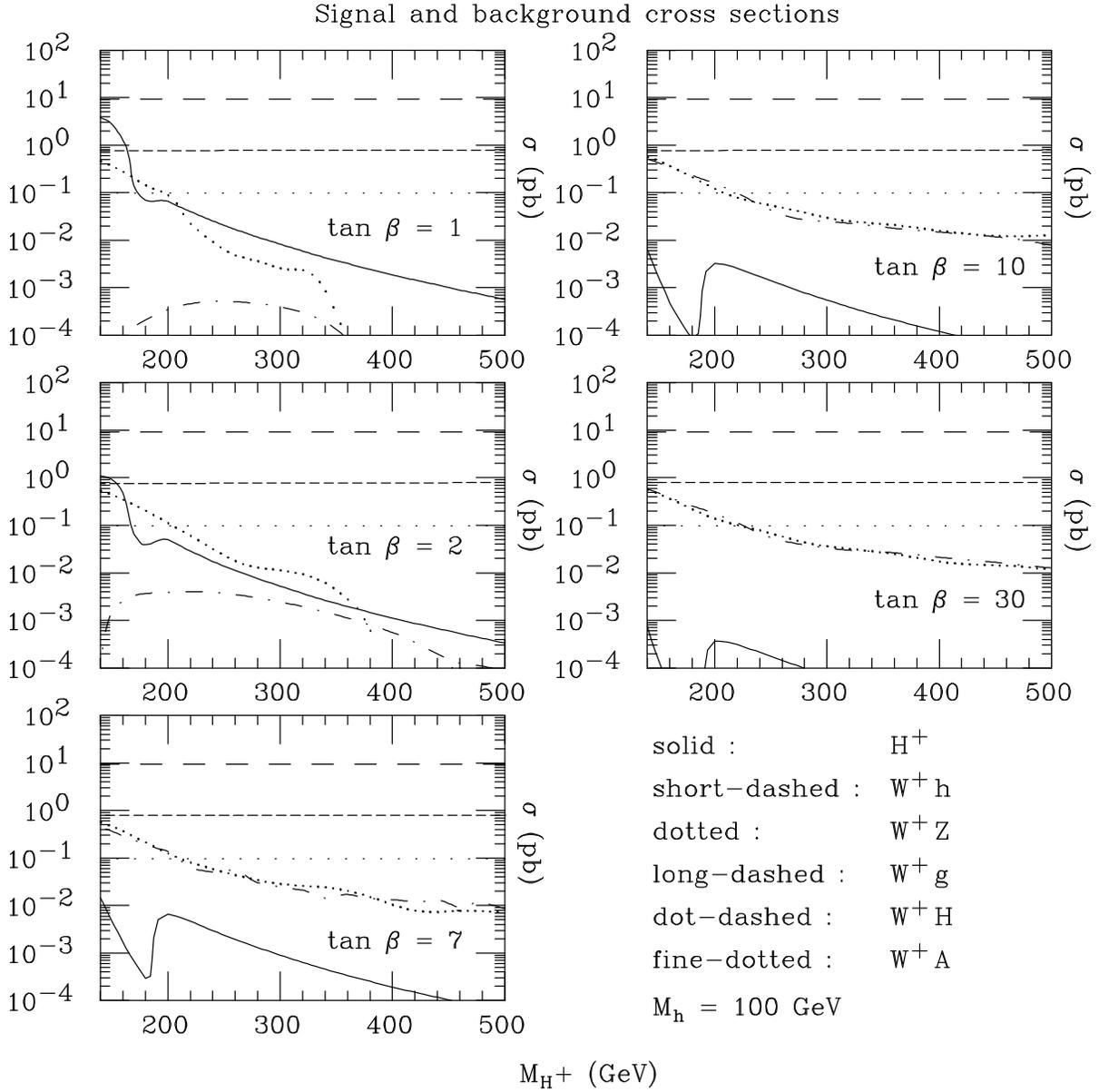}
\vspace*{0.25cm}
\caption{Same as Fig.~\ref{fig:cross90} for $M_h=100$ GeV.}
\label{fig:cross100}
\end{figure}

As a preliminary exercise, we study the total production and decay 
cross sections before any cuts, as all our reactions
are finite over their entire phase spaces (recall that $m_b\ne0$). This
is done in Figs.~\ref{fig:cross90}--\ref{fig:cross100} for the 
signal and the five background processes discussed in the 
previous Section, for five values of $\tan\beta$, over the
range 140 GeV $\Ord \MHpm\Ord$ 500 GeV, for $M_h=90$ and 100 GeV,
in the channel $X\to b\bar b$,
where $X=h,Z,g,H$ or $A$. (Of course,  the
$W^\pm Z$ and $W^\pm g$ backgrounds have no dependence on any
of the three parameters above\footnote{Note that the rates in
Figs.~\ref{fig:cross90}--\ref{fig:cross100} 
 account for the c.c. production modes as well.}.) 
As for the decay rates of the top (anti)quark and the $W^\pm$ boson, 
for sake of  simplicity, we take them equal to  $1$ for the time being. 
The signal is always dominated by the QCD background and -- at large
$\tan\beta$ -- also by the EW ones. Notice the local maxima of the
signal rates at $M_{H^\pm}\approx M_{W^\pm}+M_h$, as induced by
the opening of the $H^-\to W^-h$ decay (compare to 
Figs.~\ref{fig:BRs}--\ref{fig:BRh}), and the minima as well,
due to the onset of the $H^-\to b\bar t$ channel instead.

In the reminder of our analysis,  we assume semi-leptonic
decays of $W^+W^-$ pairs, as in Ref.~\cite{whroy}: i.e.,
$W^+W^-\to \jj~\ell^\pm\nu_\ell$ 
(hereafter, jet refers to a non-$b$-jet and $\ell=e,\mu$). However, as compared
to that analysis, we make one simplification.
Namely, we assume that {\sl one}
 top (anti)quark and the $W^\pm$ boson generated in its decay have already been
reconstructed, e.g., by using the mass selection procedure
advocated in Ref.~\cite{whroy},
either leptonically or hadronically. This allows us to greatly reduce the
complexity of our numerical calculation while -- we believe --
substantially un-affecting the relative rates of signal and backgrounds
(in fact, all processes described produce the same final state and
all involve at least one top quark).  Then we apply the following
cuts on the remaining particles (here, the label $\jet$ refers to the
decay products of the second $W^\pm$ boson present in the event, 
which can be either light-quarks or  leptons):
\begin{equation}\label{pT}
p_T(b,\jet,{\mathrm{missing}})>20~{\mathrm{GeV}}
\end{equation}
on the transverse momentum (including the missing one),
\begin{equation}\label{eta}
|\eta(b,\jet)|<2.5
\end{equation}
on the pseudorapidity, and
\begin{equation}\label{R}
\Delta R(bb,b\jet,\jet\jet)>0.4
\end{equation}
on the relative separation of $b$- and light-quark jets/leptons j, where
\begin{equation}\label{Rdef}
\Delta R({ij})=\sqrt{\Delta\eta({ij})^2+\Delta\phi({ij})^2},
\end{equation}
is defined in terms of relative differences in
pseudorapidity $\eta({ij})$ and azimuth $\phi({ij})$,
with $i\ne j=b$, j$/\ell$.
Furthermore, we impose (see also Ref.~\cite{whroy})
\begin{equation}\label{Mhcut}
|M_{bb}-M_h|<10~{\mathrm{GeV}}
\end{equation}
on exactly one pair of $b$-jets,
\begin{equation}\label{MWcut}
M_{\jet\jet}>50~{\mathrm{GeV}}
\end{equation}
on the light-jet (or lepton-neutrino) pair (recall that the $W^\pm$
can be off-shell),
and, finally, 
\begin{equation}\label{Mtcut}
|M_{bbb\jet\jet}-M_t|<20~{\mathrm{GeV}}
\end{equation}
around the top mass if three $b$'s are present
in the event (in addition to the one already used to reconstruct
the top (anti)quark). In such a case, 
one may assume that the charged Higgs boson has predominantly been 
produced in the
decay of a top (anti)quark 
(when $M_{H^\pm}\Ord m_t$). If instead only two appear, then one should
conclude that the Higgs has mainly been generated in a bremsstrahlung/fusion
process  (because
 $M_{H^\pm}\OOrd m_t$) with a $b$-(anti)quark lost along the beam pipe.
 Our $2\to3$ production mechanism naturally
allows one to emulate both
dynamics in a gauge invariant fashion,
including all interference effects. As already mentioned, however, 
we will assume
a triple $b$-tagging, this implying an overall efficiency factor
of $\epsilon_b^3$ multiplying our signal and background rates.
(Thus, the third $b$-jet in eq.~(\ref{Mtcut}) is actually 
non-$b$-tagged: it can be interpreted as the jet system
satisfying neither eq.~(\ref{Mhcut}) nor eq.(\ref{MWcut}).) 
We take $\epsilon_b=0.5$, like in \cite{whroy,whketevi} (and assume
100\% lepton identification efficiency).

\begin{figure}[p]
~\epsfig{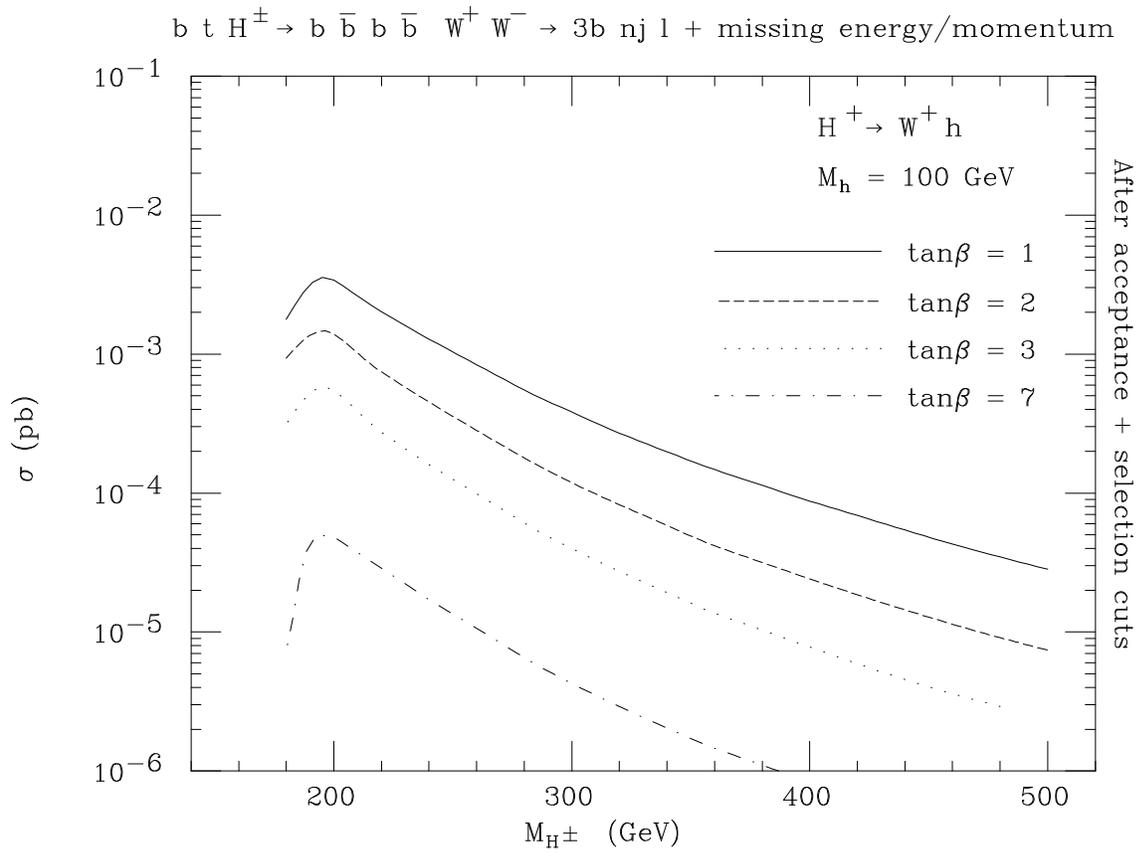}
\vspace*{0.25cm}
\caption{Cross sections of signal and backgrounds 
for selected values of $\tan\beta$ over the
mass range 160 GeV $\Ord M_{H^\pm}\Ord$ 500 GeV,
after the cuts (\ref{pT})--(\ref{Mtcut}) and including decay rates.
The mass of the lightest Higgs boson has been fixed at $M_h=100$ GeV.}
\label{fig:final}
\end{figure}

Given the signal production rates before acceptance and selection cuts,
it is clear that -- for such an $\epsilon_b$ -- 
even at high collider luminosity (i.e., $\int
{\cal L}dt=100$ fb$^{-1}$ per annum), hopes of disentangling the charged
Higgs boson of the MSSM in the $W^{\pm(*)} h$ decay channel are only confined
to the very low $\tan\beta$ region. We will thus restrict ourselves to
study  in the reminder of the paper $\tan\beta$ values which are, e.g.,
below seven.
The total signal rates after the cuts (\ref{pT})--(\ref{Mtcut})
have been applied can be found in Fig.~\ref{fig:final},
for the choices $\tan\beta=1,2,3$ and $7$, as a function of
$M_{H^\pm}$. For reference, we illustrate the `borderline' case $M_h=100$ GeV.
(Indeed, a lower $M_h$ value at $\tan\beta=2$ is in contradiction
with LEP2 data, whereas higher masses induce a far too large suppression
on BR($H^\pm\to W^{\pm*}h$): see Fig.~\ref{fig:BRh}.)
The trends in the figure are the consequence of two effects. On the
one hand,  the production cross section of $gg\to t\bar b H^-$ + c.c.
is roughly proportional to 
$(m_t^2\cot\beta^2+m_b^2\tan\beta^2)$, so that its maxima 
 occur at very low or very high $\tan\beta$. On the other hand, we have
seen how the largest $H^-\to W^{-(*)}h$ decay fraction
 is attained for $\tan\beta\approx2$. 
In the end, the largest values for $\sigma(gg\to t\bar b H^-)\times
{\mathrm{BR}}(H^-\to W^{-(*)}h)$ + c.c. are obtained for $\tan\beta=1$:
see Fig.~\ref{fig:final}. Unfortunately, such a 
$\tan\beta$ value is already excluded
in the MSSM from LEP2 data \cite{whroy}. For the optimal remaining
choice, i.e., $\tan\beta=2$, the annual rate never exceeds 140
events (before any $b$-tagging efficiency but after 
acceptance and selection cuts). The
maximum occurs at $M_{H^\pm}\approx200$ GeV, significantly above the real
threshold at $M_h+M_{W^{\pm}}\approx 180$ GeV. 

\begin{figure}[t]
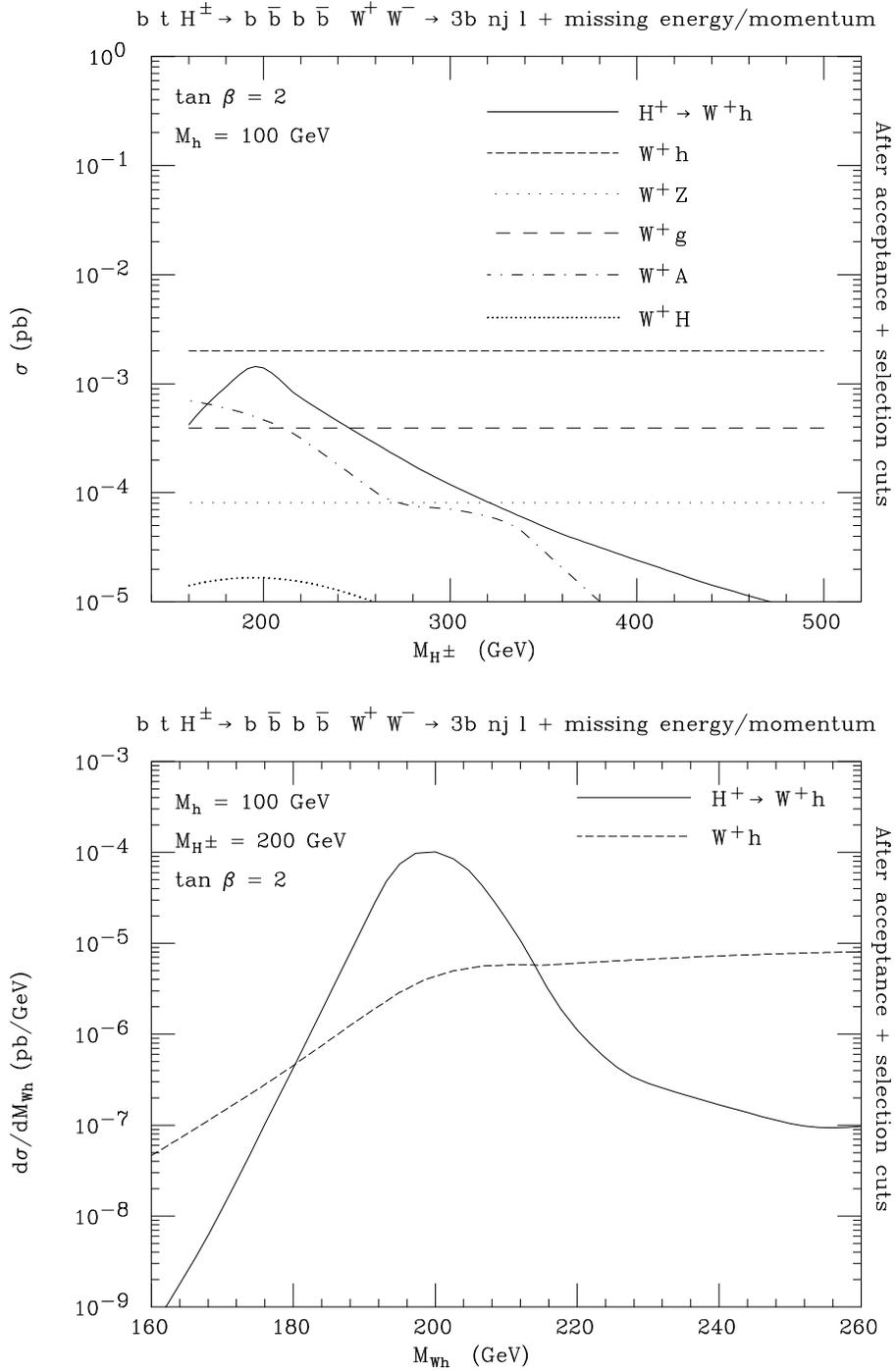

\vskip0.5cm
~\hskip1.5cm{\epsfig{file=compare.ps,height=12cm,angle=90}}
\vskip0.5cm
~\hskip1.5cm{\epsfig{file=Whmass.ps,height=12cm,angle=90}}
\vspace*{0.25cm}
\caption{(Top) Total cross sections of signal and backgrounds for
 $\tan\beta=2$ and $M_h=100$ GeV
after the cuts (\ref{pT})--(\ref{Mtcut}) and including decay rates,
 as a function of the charged Higgs boson mass
over the range 160 GeV $\Ord M_{H^\pm}\Ord$ 500 GeV. (Bottom)
Differential cross sections in 
 the reconstructed $W^\pm h$ invariant mass, for a $M_{H^\pm}=200$ GeV signal 
and for the dominant background
after the cuts (\ref{pT})--(\ref{Mtcut}) and including decay rates. 
Gaussian smearing of all transverse momenta is included:
with $(\sigma(p_T)/p_T)^2=(0.60/\sqrt{p_T})^2 + (0.04)^2$ for jets
and  $(\sigma(p_T)/p_T)^2=(0.12/\sqrt{p_T})^2 + (0.01)^2$ for leptons/missing
particles.}
\label{fig:last}
\end{figure}

We now compare such a signal with the irreducible backgrounds 1.--4., for the
same choice of $\tan\beta$ and $M_h$ (where relevant).
This is done in the upper half of Fig.~\ref{fig:last}, 
at the level of total production rates. After the cuts 
(\ref{pT})--(\ref{Mtcut}) are enforced, all background components
in 2.--4. are overwhelmed by the signal in the vicinity
of $M_{H^\pm}=200$ GeV, whereas  the $W^\pm h$
continuum production is always larger than the 
$H^\pm\to W^\pm h$ resonant channel. Thus, it is relevant to compare
the last two processes in the `reconstructed'
invariant mass $M_{W^\pm h}$, i.e., that obtained
from pairing the two $b$-jets fulfilling condition (\ref{Mhcut})
and the two
light-quark jets (or, alternatively, the lepton-neutrino pair)
satisfying eq. (\ref{MWcut}) and not already reconstructing $M_{W^\pm}$
on their own and $m_t$ in association with any of the $b$'s (see
Ref.~\cite{whroy}). The spectrum in this variable is presented
in the lower half of  Fig.~\ref{fig:last}, for our ideal case 
$M_{H^\pm}=200$ GeV (and, again, $\tan\beta=2$ and $M_h=100$ GeV).
For such MSSM parameter combination, the charged Higgs signal is
well above the continuum for values of $M_{W^\pm h}$ which are
$\pm20$ GeV from $M_{H^\pm}$.
(To vary $M_{H^\pm}$ and/or $\tan\beta$
basically corresponds to rescale the solid line in the last plot
by a constant factor, according to the rates in Fig.~\ref{fig:final}.)

For reference,
Tab.~\ref{tab:cuts} presents the number of events of resonant and
continuum $W^\pm h$ production at the LHC, after 300 inverse
femtobarns of collected luminosity, for $\epsilon_b^3=0.125$, in the window
 $|M_{H^\pm}-M_{W^\pm h}|<$ 40 GeV, for the three values
$M_{H^\pm}=180,200$ and 220 GeV. Although very small, a $H^-\to W^- h$
signal is generally observable above the $W^- h$ continuum for
$M_{H^\pm}$ around $200$ GeV. Our numbers are roughly consistent with
those in Ref.~\cite{whroy}, if one considers that we neglect
the finite efficiency of reconstructing one $W^\pm$ boson and the associated
top (anti)quark and since we have chosen somewhat different cuts.
Therefore, in the end, the dominant backgrounds remain (in the $3b$-tagged
channel) the $H^- b\bar t$ + c.c. decay and the QCD noise involving
misidentified gluons, i.e., those already identified in Ref.~\cite{whroy}.

\section*{4. Conclusions}

In summary, in this paper, we have complemented a
previous analysis \cite{whroy} of the production
and decay of charged Higgs bosons of the MSSM at the LHC, in the channels
$gg\to t\bar b H^-$ and $H^-\to W^{-(*)}h$ (and charged conjugated modes),
respectively, by considering several irreducible backgrounds in the 
$3b$-tagged channel,
i.e., $t\bar b H^-$ + c.c. $\to 3b~{\mathrm{2~jets}}~\ell$ + `missing energy'
(where the initial $b$-(anti)quark is usually
 lost along the beam pipe), which had not yet been considered.

We have found that, after standard acceptance cuts and a kinematic
selection along the lines of the one outlined in Ref.~\cite{whroy},
the dominant background among those considered here is the continuum production
$gg\to t\bar b  W^{-(*)}h$ + c.c. However, the latter has been found to
lie significantly below the signal in the only region where this
 is detectable: when $\tan\beta\approx2-3$ and 
$M_{H^\pm}\approx 200$ GeV (with $M_h$ around 100 GeV, close to the 
latest LEP2 constraints). Thus, the chances of detecting the
$H^-\to W^{-}h\to W^-b\bar b$ decay in such a (narrow) 
region of the MSSM parameter space depend mainly on the
interplay between this mode, the competing one $H^-\to b\bar t\to
W^-b\bar b$ and the QCD background with mistagged gluons, as are the latter 
two that clearly
overwhelm the former (recall the last figure in \cite{whroy}).

We have carried out our analysis at parton level, without showering
and hadronisation effects but emulating typical detector smearing.
We are confident that its salient features should survive a more
sophisticated simulation, 
such as the one presented in Ref.~\cite{whketevi}. Besides,
our results concerning the backgrounds can be transposed to the case
of non-minimal SUSY models (where
the $H^\pm$ discovery potential can extend to a much larger
portion  of parameter space), such as those considered in 
Ref.~\cite{whroy},
so that also in these scenarios the irreducible backgrounds analysed
here can be brought under control.

\subsection*{Acknowledgements}

The author is grateful to the UK-PPARC for financial support. Furthermore,
he thanks D.P. Roy for his remarks, which induced him to eventually
considering the subject of this research. He also thanks
K.A. Assamagan for several useful discussions.
Finally,
many conversations with K. Odagiri are  acknowledged, as well
as many numerical comparisons against and the use of some of his programs. 

\goodbreak


\vskip6.0cm

\begin{table}[h!]
\begin{center}
\begin{tabular}{|c|c|c|c|}
\hline
\multicolumn{4}{|c|}
{Number of events after 300 fb$^{-1}$ (including c.c. channels)}\\
\hline
\hline
\multicolumn{4}{|c|}
{$\tan\beta=2$
\qquad\qquad
\qquad\qquad
\qquad\qquad
$M_h=100$ GeV}\\
\hline\hline
$M_{H^\pm}$ (GeV) & 
$t\bar b H^-$     & 
$t\bar b W^-h$    & 
$S/\sqrt{B}$      \\
\hline
$180$ & 
$34$ & 
$6$ & 
$13$ \\
$200$ & 
$52$ & 
$11$ & 
$16$ \\
$220$ & 
$27$ & 
$17$ & 
$7$ \\
\hline\hline
\multicolumn{4}{|c|}
{MRS-LO(05A)} \\
\hline\hline
\multicolumn{4}{|c|}
{$3b$-tag
\qquad\qquad
\qquad\qquad
\qquad\qquad
\qquad\qquad
All cuts}
\\ \hline
\end{tabular}
\end{center}
\caption{Number of signal ($S$, $t\bar b H^-$) and
dominant background ($B$, $t\bar b W^-h$) events,
along with the statistical significance $S/\sqrt B$,  
after  the implementation of the
cuts (\ref{pT})--(\ref{Mtcut}) in the decay channel 
$W^+W^- h\to \jj~\ell^\pm\nu_\ell~b\bar b$. Rates
are given for $\tan\beta=2$, $M_h=100$ GeV, three
choices of $M_{H^\pm}$,
as obtained by using the MRS-LO(05A) set of PDFs, after 
300 fb$^{-1}$ of luminosity and for $\epsilon_b=0.5$.
}
\label{tab:cuts}
\end{table}

\clearpage\thispagestyle{empty}

\clearpage\thispagestyle{empty}

\clearpage\thispagestyle{empty}

\clearpage\thispagestyle{empty}

\clearpage\thispagestyle{empty}

\clearpage\thispagestyle{empty}


\begin{thebibliography}{99}


\bibitem{HHG} J.F.~Gunion, H.E.~Haber, G.L.~Kane and S.~Dawson,
                ``The Higgs Hunter Guide''
                (Addison-Wesley, Reading MA, 1990).

\bibitem{CMS} CMS Technical Proposal, CERN/LHC/94-43 LHCC/P1, December 1994.

\bibitem{ATLAS} ATLAS Technical Proposal,
CERN/LHC/94-43 LHCC/P2, December 1994.

\bibitem{FNAL} 
CDF Collaboration, {\it Phys. Rev. Lett.}   {\bf 79} (1997) 357;
D0  Collaboration, \prl   82 1999 4975.

\bibitem{bg} J.F. Gunion, H.E. Haber, F.E. Paige, W.-K. Tung and
S.S.D. Willenbrock, \np B294 1987 621.

\bibitem{bq} S. Moretti and K. Odagiri, \pr D55 1997 5627.

\bibitem{BBK} A.A. Barrientos Bendez\'u and B.A. Kniehl,
{\it Phys. Rev.} {\bf D59} (1999) 015009.

\bibitem{ioekosuke} S. Moretti and K. Odagiri,
{\it Phys. Rev.} {\bf D59} (1999) 055008.

\bibitem{roger} V. Barger, R.J.N. Phillips and D.P. Roy,
\pl B324 1994 236.

\bibitem{gunion} J.F. Gunion and S. Geer, preprint UCD-93-32,
September 1993, {\tt hep-ph/9310333};
J.F. Gunion, \pl B322 1994 125.

\bibitem{roy} D.J. Miller, S. Moretti, D.P. Roy and W.J. Stirling,
{\it Phys. Rev.} {\bf D61} (2000) 055011.

\bibitem{roy1}  S. Moretti and D.P. Roy,
{\it Phys. Lett.} {\bf B470} (1999) 209.


\bibitem{kosuke} K. Odagiri, preprint RAL-TR-1999-012, February 1999,
{\tt hep-ph/9901432}.

\bibitem{ray} S. Raychaudhuri and D.P. Roy,
{\it Phys. Rev.} {\bf D53} (1996) 4902.

\bibitem{Ben} B.K.  Bullock,  K. Hagiwara and A.D.  Martin, 
{\it Phys. Rev. Lett.} {\bf 67} (1991) 3055; {\it Nucl. Phys.}
{\bf B395} (1993) 499.

\bibitem{newroy} D.P. Roy {\it Phys. Lett.} {\bf B459} (1999) 607.

\bibitem{work} K.A. Assamagan, ATLAS Internal Note ATL-PHYS-99-013 (1999);
K.A. Assamagan, A. Djouadi,
M. Drees, M. Guchait, R. Kinnunen,
J.L. Kneur, D.J. Miller, S. Moretti,
K. Odagiri and D.P. Roy,
contribution to the Workshop `Physics at TeV Colliders',
Les Houches, France, 8-18 June 1999,
{\tt hep-ph/0002258}  (to appear in the proceedings).

\bibitem{BR1} S. Moretti and W.J. Stirling, {\it Phys. Lett.} {\bf 
B347} (1995)
291; Erratum, {\it ibidem} {\bf B366} (1996) 451.

\bibitem{BR2} E. Barradas, J.L. Diaz-Cruz, A. Gutierrez and A. Rosado,
{\it Phys. Rev.} {\bf D53} (1996) 1678;
A. Djouadi, J. Kalinowski and P.M. Zerwas, {\it Z. Phys.} {\bf C70} (1996)
435; 
E. Ma, D.P. Roy and J. Wudka, 
{\it Phys. Rev. Lett.} {\bf 80} (1998) 1162.

\bibitem{whroy} M. Drees, M. Guchait and D.P. Roy,
\pl B471 1999 39. 

\bibitem{whketevi} K.A. Assamagan, ATLAS Communication ATL-PHYS-99-025
(1999); K.A. Assamagan, A. Djouadi,
M. Drees, M. Guchait, R. Kinnunen,
J.L. Kneur, D.J. Miller, S. Moretti,
K. Odagiri and D.P. Roy, in Ref.~\cite{work}.


\bibitem{mh} See, e.g.:
LEP Higgs Working Group, http://www.cern.ch/LEPHIGGS/.

\bibitem{gg} J.L. Diaz-Cruz and O.A. Sampayo, {\it Phys. Rev.} 
{\bf D50} (1994) 6820;
F. Borzumati, J.L. Kneur and N. Polonsky, 
{\it Phys. Rev.} {\bf D60} (1999) 115011.

\bibitem{KS} R.~Kleiss and W.J.~Stirling, \np {B262} {1985} {235}.

\bibitem{berends} F.A. Berends, P.H. Daverveldt and R. Kleiss,
\np {B253} {1985} {441}.

\bibitem{mana} C.~Mana and M.~Martinez,
\np {B287} {1987} {601}.

\bibitem{ioPRD} S. Moretti, {\it Phys. Rev.} {\bf D50} (1994) 2016.

\bibitem{VEGAS} G.P.~Lepage, {\it Jour. Comp. Phys.} {\bf 27} (1978) 192.

\bibitem{RAMBO} R. Kleiss, W.J. Stirling and S.D. Ellis,
{\it Comput. Phys. Commun.} {\bf 40} (1986) 359.

\bibitem{hamid} 
 H. Kharraziha and S. Moretti, preprint
 DESY 99-133, RAL-TR-1999-061, TSL/ISV-99-0216,
September 1999, {\tt hep-ph/9909313}.

\bibitem{MRS98LO} A.D. Martin, R.G. Roberts, W.J. Stirling and R.S. Thorne,
{\it Phys. Lett.} {\bf B443} (1998) 301.

\bibitem{carena}
H.E. Haber, R. Hempfling and A.H. Hoang, {\it Z. Phys.} {\bf C75}
(1997) 539; M. Carena, J. Espinosa, M. Quiros and C. Wagner,
{\it Phys. Lett.} {\bf B355} (1995) 209.

\end{thebibliography}
\end{document}